\documentclass[pra,aps,reprint,showpacs,amsmath,amssymb,groupedaddress,longbibliography]{revtex4-1}
\usepackage{hyperref}
\usepackage{epsfig}
\usepackage{color}
\usepackage{graphicx}
\usepackage{bm}
\usepackage{epstopdf}
\usepackage{amsmath}
\usepackage{tikz}
\usepackage{todonotes}

\usepackage{general_latex}

\begin{document}

\title{Self-consistent determination of the many-body state of ultracold bosonic atoms \\ in a one-dimensional harmonic trap}
\author{Oleksandr V. Marchukov}
\affiliation{School of Electrical Engineering, Faculty of Engineering, Tel Aviv University, 6997801, Tel Aviv, Israel}
\author{Uwe R. Fischer}
\affiliation{Department of Physics and Astronomy, Seoul National University, 
Center for Theoretical Physics, 
08826 Seoul, Republic of Korea}

\begin{abstract}
We study zero-temperature quantum 
fluctuations in harmonically trapped one-dimensional interacting Bose gases, using the 
self-consistent multiconfigurational time-dependent Hartree method. 
We define {\em phase fluctuations} 
from the full single-particle density matrix by the spatial decay exponent of off-diagonal long-range order.  
In a regime of mesoscopic particle numbers and moderate contact couplings,  
we derive the spatial dependence of the amplitude of phase fluctuations, determined from the {\em self-consistently} derived shape of the field operator orbitals and Fock space orbital occupation amplitudes. It is shown that the phase fluctuations display a peak, which in turn corresponds to a dip of the first-order correlations in position space, akin to what has previously been obtained 
in the Tonks-Girardeau limit of very large interactions and low densities. 
\end{abstract}

\date{\today}

\maketitle

\section{Introduction}
The standard paradigm of the weakly interacting Bose gas is Bogoliubov theory \cite{bogoliubov}.
A single orbital is occupied macroscopically, with a continuum of 
excitations existing on top of this {\em condensate}.
On the other hand, in spatial dimension smaller than three, the absence of such a Bose-Einstein condensate 
(BEC) in infinitely extended homogeneous systems directly follows from the fundamental quantum statistical Bogoliubov inequality \cite{bogoliubov}, as derived in \cite{mermin1966, hohenberg1967,Pitaevskii1991}. 
Recently, the advances in modern precision experiments using ultracold atomic gases have 
rekindled the interest in the coherence properties of low-dimensional quantum gases 
\cite{pethick2008,cazalilla2011}.

For finite systems, which are those actually realized in experiments with trapped Bose gases, 
the question of the existence of BEC in low dimensions is more intricate. 
For example, a one-dimensional (1D) BEC, at given interaction coupling and density,  
can exist up to a critical length of the condensate \cite{fischer2002,Fischer2005}.
Within the realm of Bogoliubov theory \cite{petrov2000,petrov2001}, 
its extension to quasicondensates \cite{mora2003}, and the Luttinger liquid approach \cite{Gangardt2003},  
phase fluctuations have been shown to gradually destroy off-diagonal-long-range-order (ODLRO) \cite{Yang}, in finite 1D Bose-Einstein condensates, and to lead to a characteristic power law decay of 
correlation functions. 
These {\em phase-fluctuating condensates} have been probed in numerous experiments, initially in   
\cite{dettmer2001,hellweg2003,Aspect}, and with increasing sophistication 
in recent years cf., e.g.,  \cite{hofferberth,manz,schweigler2015,jacqmin,fang}.

We employ in what follows a fully self-consistent 
many-body approach 
to describe phase-fluctuating 1Dcondensates,   
the multiconfigurational time-dependent Hartree (MCTDH) method for bosons~\cite{Meyer1990,alon2008}. 
The latter 
accommodates the correlations between all significantly occupied orbitals.  
As a consequence, the 
phase fluctuations develop a peak in position space, which is akin to what is obtained for a pair of bosons in Monte Carlo simulations at (infinitely) large interactions 
and low densities, that is in the 
Tonks-Girardeau limit \cite{Minguzzi}.
We demonstrate that the fluctuation peak height is essentially proportional to the {\em degree of  fragmentation} (defined in the below as the relative occupation of states outside the most occupied one), and are hence establishing the connection of  the maximal magnitude of phase fluctuations (which are nonlocal in space)  and the fragmentation degree.  

{The importance of many-body correlations between all modes in the phase-fluctuating regime which we demonstrate can be revealed by realistically implementable experiments cf., e.g., \cite{hofferberth,manz,schweigler2015,jacqmin,fang}.  
This relative facility of experimental access} is in marked contrast to the large degrees of fragmentation necessary to observe 
significant density-density correlations \cite{Kang}.  
Our self-consistent many-body results therefore pave the way to study experimentally 
the low-dimensional many-body phenomena related to condensate fragmentation 
in a single harmonic trap 
\cite{Bader,FischerBader}. 
They provide a benchmark to correlate theory and experiment in quantum many-body physics, and in principle to arbitrarily high order in the correlation functions \cite{schweigler2015}. 

{Importantly, before we describe in detail our formalism and results, we note that we consider bosons in a {\em single} 1D harmonic trap, not in a 1D double well trap, for which the MCTDH approach has been employed previously, cf., e.g., \cite{Zoellner1,Sakmann}.
In the latter case, the {\em externally imposed} one-body potential with a double-minimum preimposes a structure of the orbitals quite distinct from that of a single trap, and the fragmentation is (to some extent) extrinsically engineered as opposed to intrinsically caused by interactions. }

{Furthermore, a salient property of the MCTDH method is to allow the basis functions to change their shape (as described in Appendix A of the paper), in contrast to exact diagonalization methods using a fixed basis of harmonic oscillator eigenfunction, cf., e.g., Ref~\cite{Haugset}.
The MCTDH method is therefore capable to self-consistently describe spatial correlation dips in the first-order correlation 
function, as we reveal below.}


\section{Many-Body Formalism}
In the interacting Bose gas, the (basis invariant) definition of BEC is due to Penrose and Onsager~\cite{penrose1956}. 
It employs the position space single-particle density matrix in its eigenbasis, 
\begin{equation}
\rho^{(1)}(x, x') \coloneqq \left\langle \hat {\psi^{\dagger}}(x) \hat \psi(x') \right\rangle = \sum_{i=1}^M N_i \varphi^{\ast}_i(x) \varphi_i(x'), \label{SPDM} 
\end{equation} 
where $\hat {\psi^{\dagger}}(x)$ and $\hat{\psi}(x)$ are bosonic creation and annihilation
field operators, respectively. 
The angled brackets indicate quantum-statistical average over states; we work at zero temperature.
Furthermore, $\varphi_i(x)$ are the single-particle wavefunctions (called \textit{natural orbitals} in this eigenbasis of $\rho^{(1)}$), $N_i$ are their occupation numbers, and $M$ is the number of orbitals (in practice fixed 
by the available computational resources). 
The definition \cite{penrose1956} states that if a subset of the eigenvalues $N_i$ are ``macroscopic,'' 
i.e. some of the $n_i=N_i/N$ remain finite in the large $N$ limit,
then the many-body state of the Bose gas  is a simple or fragmented BEC 
when the cardinality of this subset is one or larger than one, 
respectively \cite{leggett2003,mueller2006}. We note that this formal definition 
of fragmentation cannot be rigorously applied in our case (given that both $N$ and the $N_{i\neq 1}$ 
are relatively small in the investigated range of parameters, see below). We however denote in the following the quantity 
$1-n_1$,  
 where $n_1=N_1/N$ is the  relative occupation number of the energetically lowest orbital, 
as the ``degree of fragmentation,''  
in the sense of a 
shorthand for the relative number of particles residing in all orbitals which are outside the condensate
(also taking into account, when using this term, the fact that the field operator truncation is at finite $M$). 

The Hamiltonian that we consider to address the many-body problem describes bosons interacting by a
contact pseudopotential and placed in a harmonic trap:
\begin{equation}
\label{hamiltonian}
 H = \sum_{i=1}^{N} \left (\frac{p_i^2}{2m} + \frac{1}{2}m\omega^2 x_i^2 \right ) + g_{\rm 1D}\sum_{i>j} \delta(x_i - x_j).
\end{equation}
Here, $x_i$ and $p_i$ are 1D position and momentum operators of a given atom (or molecule) $i$, respectively, $\omega$ is the frequency of the trapping harmonic potential, $N$ is the number of particles, $m$ is their mass,
and $g_{\rm 1D}$ is the contact coupling. In order to establish the connection with experiment, 
one considers a quasi-1D gas, {i.e.} the particles 
are trapped in a three-dimensional (3D) harmonic potential that is strongly anisotropic, with the transverse
frequency being much larger than the axial one, $\omega / \omega_\perp\ll1$, 
such that transverse motion is confined (frozen) to the ground state. 
The 1D coupling strength, $g_{\rm 1D}$, is related to the 3D scattering length, $a_{\rm sc}$, by 
$g_{\rm 1D} = 4\pi \hbar^2 a_{\rm sc}/(\pi m l_\perp^2)$ far away from geometric scattering resonances, 
where $l_\perp=\sqrt{\hbar/m\omega_\perp}$ is the transverse oscillator length, {when one is far away from the geometric scattering resonance which occurs for $a_{\rm sc} \sim l_\perp$} ~\cite{olshanii1998}.

The homogeneous quantum many-body problem corresponding to the Hamiltonian \eqref{hamiltonian}  
without harmonic trap ($\omega=0$) can be solved
by Bethe ansatz, as was shown long ago by Lieb and Liniger~\cite{lieb1963a}. 
On the other hand, for trapped (spatially confined) and thus generally inhomogeneous systems, the mathematical case most relevant to actual experiments,  the exact solution is not known  (except for the hard-core limit 
\cite{volosniev2014a}).
There are several numerical methods that are applicable to the present problem, such as, for example, 
density matrix renormalization group~\cite{white1992, hallberg2006} and quantum Monte Carlo methods~\cite{austin2012}.
Here, we used the multiconfigurational time-dependent Hartree (MCTDH) method for bosons~\cite{alon2008} and, in particular, its implementation MCTDH-X~\cite{mctdhx,LodeMulti}. This powerful method, long known in physical chemistry for distinguishable particles \cite{Meyer1990,meyer2009}, has since the advent of ultracold quantum gases  
proven its value for the study of the correlation properties of fragmented BECs cf., e.g.,~\cite{sakmann2009,lode2012,streltsov2013,fischer2015,klaiman2015,klaiman2016,tsatsos2016a,sakmann2016,LodeBruder}. 
The MCTDH method constructs the many-body wavefunction as a sum of all possible configurations of distributing $N$ particles over $M$ time-dependent single-particle wavefunctions (orbitals). 
The system of equations for both Fock space occupation distribution amplitudes and shape of orbitals, 
obtained after applying the Dirac-Frenkel variational principle, is solved self-consistently and yields the many-body wavefunction of the system, cf.\,Appendix A for a concise summary.
{Numerous MCTDH studies have demonstrated the importance of self-consistency, and even 
when energy and density differences to a non-self-consistent mean-field treatment are very small. See for an illustration, e.g., Ref.\,\cite{klaiman2015}, which shows that even in a 
large $N$ limit in which energy and density are described 
exactly by the mean field theory, and the occupation of higher orbitals becomes negligibly small,
subtle correlations are revealed by the self-consistent approach.
 

\section{Definition of Phase Fluctuations}
Using the representation of the field operators 
\bea \hat \psi (x) = e^{i\hat \phi(x)}\sqrt{\hat \rho(x)},\quad \hat {\psi}^\dagger(x) = \sqrt{\hat \rho(x)} e^{-i\hat \phi(x)},
\ea 
 where $\hat \rho(x)$ is the particle density operator and $\hat \phi(x)$
is a (hermitian) phase operator, the single-particle density matrix can be written in the form 
\bea
\rho^{(1)}(x, x') 
\label{def_matrix} 
= \left\langle \sqrt{\hat \rho(x)} e^{-i(\hat \phi(x) - \hat \phi(x'))} \sqrt{\hat \rho(x')}\right\rangle.\ea 
It is well known that one should exercise 
care when defining the phase operator in this way, 
as thoroughly reviewed in Refs.~\cite{carruthers1968, lynch1995}; also see the coarse-graining procedure applied in \cite{mora2003}.
We will use the definition above which coincides with the traditional Dirac approach~\cite{lynch1995}. 

{We now proceed to our {\em definition of phase fluctuations.} 
{To establish our notion of phase fluctuations from the full single-particle density matrix, 
we resort
for defining phase fluctuations  to a definition akin to that employed in \cite{ho1999, petrov2000}.
We posit, replacing in \eqref{def_matrix} 
the local density operator by its mean, the relation} 
\begin{equation}
\label{dens_matr}
 \sqrt{\rho(x) \rho(x')} 
 \exp\left[-\frac{1}{2}\left\langle \hat{\delta\phi}_{xx'}^2 \right\rangle\right]
 \coloneqq \rho^{(1)}(x, x') ,
\end{equation}
where $\hat {\delta \phi}_{xx'} = \hat \phi(x) - \hat \phi(x')$ is the phase difference operator and 
$\rho(x)=\langle  \hat {\psi^{\dagger}}(x) \hat \psi(x) \rangle$. 
{Note that there is no need to  justify the apparent ``neglect" of density fluctuations when writing \eqref{dens_matr}, because beyond mean-field, when one solves the many-body problem with MCTDH, there is no unique phase reference belonging to a single macroscopically occupied orbital as in Bogoliubov theory:  
The notion of a phase fluctuation, when several macroscopically occupied orbitals are present,   
must by necessity remain a derived concept, namely solely {\em by definition} from the full single-particle density matrix. 
Defined in this way, the generalized, effective phase fluctuations simply  
represent (twice) the decay exponent of ODLRO \cite{Yang}.}

Solving \eqref{dens_matr} for $\langle \hat{\delta\phi}_{xx'}^2 \rangle$, we obtain a direct relation of mean-square phase fluctuations and single-particle 
density matrix in position space:
\begin{equation}
\label{phase_fluct1}
 \langle \hat {\delta \phi}^2_{xx'} \rangle = -2 \ln{\left [ \frac{\rho^{(1)}(x, x') }
{\sqrt{\rho(x)\rho(x')}}\right]}.
\end{equation}
Using the diagonalized form of the single-particle density matrix in Eq.\,\eqref{SPDM} 
one can rewrite Eq.~\eqref{phase_fluct1} as
\begin{equation}
\label{ph_fluct}
 \langle \hat {\delta \phi}^2_{xx'} \rangle = -2 \ln{\left [ \frac{ \sum_{i=1}^M N_i \varphi^{\ast}_i(x') \varphi_i(x)}
{\sqrt{\rho(x)\rho(x')}}\right ]}.
\end{equation}
We evaluate \eqref{ph_fluct} after finding the many-body ground state of the system using MCTDH-X \cite{mctdhx}. 
For our calculations with $N=10$ and $N=30$,   
we used five available orbitals, $M = 5$, to guarantee that there is no significant 
occupation in the highest orbitals. To ensure convergence, we compared the $M=5$ results for the occupation numbers with those obtained by using a much larger $M=10$ basis, and found that the relative error is not higher than $\approx0.5\,\%$ for the strongest interaction considered, i.e.  
$[n_1(M=10) - n_1(M=5)] / n_1 (M=10) \approx 0.005$ for $g_0 = 1.0$.
Table~\ref{tableError} provides more detailed information on relative errors of 
energy and occupation number for $N=10$ and varying values of the interaction strength. 
We conclude that even a very moderate size of basis $M=5$ gives rather reliable results. 
For a larger particle number, $N=100$, we used four available orbitals, $M=4$, 
to remain within reasonable time constraints. Nevertheless, the highest orbital, also when employing this constraint to smaller $M$, 
is not significantly occupied. See for a detailed discussion of the $N=30$ and $N=100$ results Appendix B.

\begin{table}
\begin{tabular}{| l | l | l | l | l | l |}
 \hline
 $g_0$&$\Delta_{\rm rel}E$ & $\Delta_{\rm rel}n_1$ & $\Delta_{\rm rel}n_2$ \\ \hline \hline
 $0.1$  & $-6.8\times10^{-4}$ & $-6.3\times10^{-5}$ & $2.0\times10^{-2}$ \\ \hline
 $0.5$  & $-8.2\times10^{-3}$ & $-1.3\times10^{-3}$ & $1.9\times10^{-2}$ \\ \hline
 $0.75$ & $-1.4\times10^{-2}$ & $-3.1\times10^{-3}$ & $2.3\times10^{-2}$ \\ \hline
 $1.0$  & $-2.1\times10^{-2}$ & $-5.8\times10^{-3}$ & $2.9\times10^{-2}$ \\ \hline
 \end{tabular}
 \caption{Table of relative approximation errors, $\Delta_{\rm rel} f = (f_{M=10} - f_{M=5}) / f_{M=10}$, calculated using basis sizes $M=5$ and $M=10$ for energy and occupations $n_1$ and $n_2$,
 $f\coloneqq \{E,n_1,n_2\}$, with $N=10$ and various 
couplings $g_0$ [Eq.\,\eqref{g0def}].}
\label{tableError}
\end{table}
It is convenient to introduce a dimensionless, scaled interaction parameter 
\begin{equation}
g_0 = mg_{\rm 1D}l/\hbar^2, 
\label{g0def}
\end{equation}
 where $l = \sqrt{\hbar/m\omega}$ is the axial harmonic oscillator length.  
We considered the range $g_0 = 0.1$\,--\,$1.0$, which even for our relatively small number of particles 
corresponds to the weakly-interacting Bose gas in the Thomas-Fermi regime~\cite{petrov2000}. This range of couplings 
is well within current experimental possibilities~\cite{hofferberth,manz,schweigler2015,fang,jacqmin}.
This regime may be further characterized by using a dimensionless Hartree parameter~\cite{gudyma2015}, 
which in our units reads $\lambda = \frac{2}{g_0 N}$. For the majority of the parameter values that we employ in our calculations we have $\lambda < 1$, and are therefore in a 
Thomas-Fermi type regime according to the terminology employed in \cite{gudyma2015}. 
{In Fig.~\ref{densities}, we plot the densities for $N=10$ particles and varying interaction strength $g_0$, with the values displayed in tables \ref{tableError} and \ref{table10}. 
Except for $g_0 = 0.1$ and hence $\lambda=2$, indeed we see that the density profiles are close to the shape of inverted parabolas. For $g_0 = 0.1$, the profile is closer to a Gaussian distribution, in accordance with Ref.~\cite{gudyma2015}.

{Alternatively, to further aid understanding of the regime we work in,  
one may define an effective Lieb-Liniger parameter \cite{lieb1963a} 
by $\gamma_{\rm LL}=\frac{m}{\hbar^2}\, g_{\rm1D}/ \bar\rho = g_0 \frac{2R}{N l}$ , where $2R$ is a suitably defined length of the gas in the harmonic trap, cf.~Fig.~\ref{densities}.   
We have, for the majority of our parameters, $\gamma_{\rm LL} \ll 1$, and the largest value ($g_0=1,N=10,2R/l \simeq 6$) is $\gamma_{\rm LL} \simeq 0.6$, so we are throughout in a weakly to maximally intermediate coupling regime, as required for the number of orbitals $M$ we can calculate with. 
}}
\begin{figure}[t]
\hspace*{-2em}
 \includegraphics[width=0.35\textwidth]{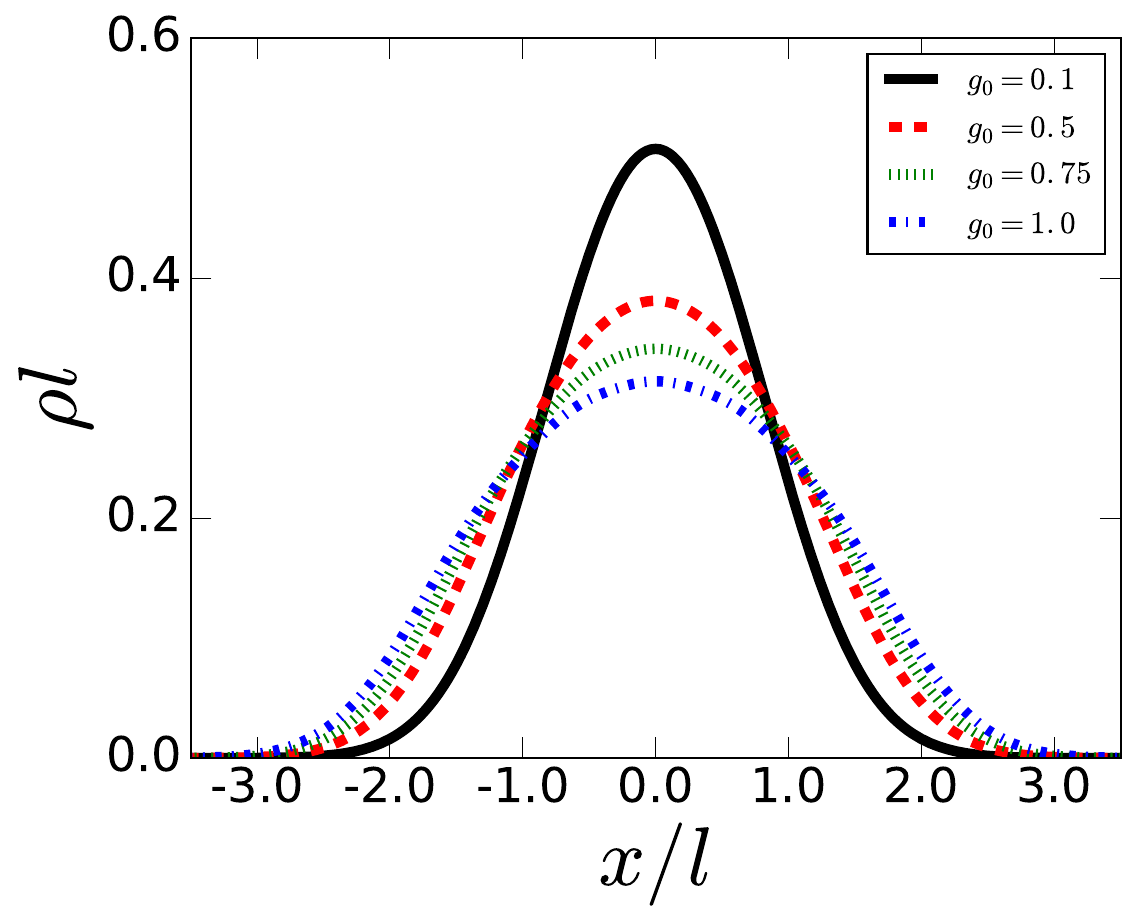}
 \caption{Density profiles $\rho(x)$ for $g_0=0.1$ (black solid), $0.5$ (red dashed), $0.75$ (green dotted), and $1.0$ (blue dashed dotted) and $N=10$ atoms in the trap.}
  \label{densities}
\end{figure}
\section{Results}
\begin{table}[b]
\begin{tabular}{| l | l | l | l | l | l |}
 \hline
 $g_0$&$n_1$ & $n_2$ & $n_3$ & $n_4$ & $n_5$ \\ \hline \hline
 $0.1$  & $0.99875$ & $0.00095$ & $0.00022$ & $0.00004$ & $0.00002$ \\ \hline
 $0.5$  & $0.98068$ & $0.01338$ & $0.00425$ & $0.00119$ & $0.00049$ \\ \hline
 $0.75$ & $0.96611$ & $0.02241$ & $0.00788$ & $0.00254$ & $0.00105$ \\ \hline
 $1.0$  & $0.95152$ & $0.03087$ & $0.01168$ & $0.00418$ & $0.00175$ \\ \hline
 \end{tabular}
 \caption{Table of relative occupation numbers, $n_i=N_i/N$, for $N=10$ and various 
couplings $g_0$ [Eq.\,\eqref{g0def}].}
\label{table10}
\end{table}
\begin{figure}
   \includegraphics[width=0.42\textwidth]{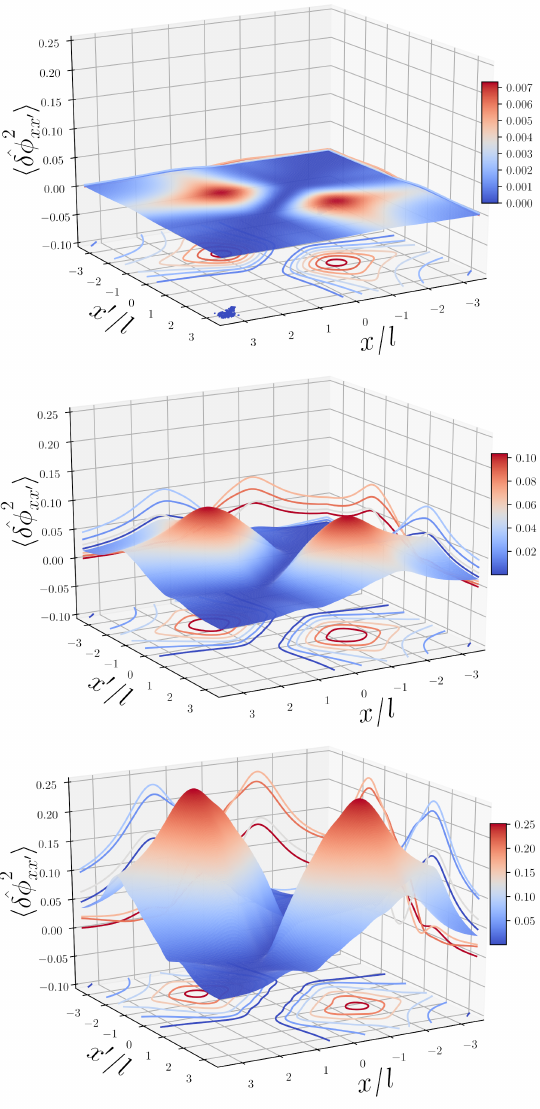}
 \caption{Mean-square quantum phase fluctuations $\langle \hat{\delta \phi}^2_{xx'} \rangle$ for $N=10$ and increasing coupling strength:  
 Top $g_0 = 0.1$, Middle $g_0 = 0.5$, and Bottom $g_0 = 1.0$. The maxima along the off-diagonals, $x'=-x$, correspond to the fact that the gas  
 becomes phase-uncorrelated in distant regions of the cloud.}
  \label{surf_fluct}
\end{figure}
\begin{figure}[t]
 \includegraphics[width=0.38\textwidth]{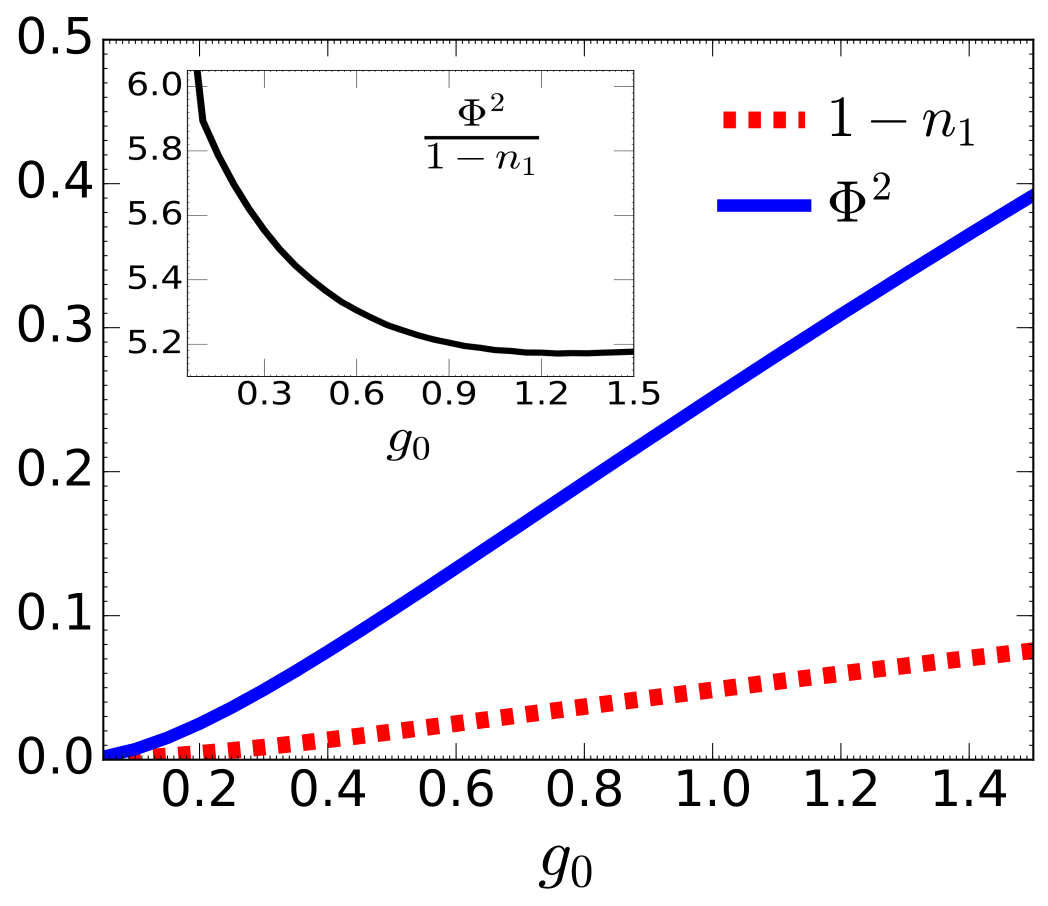}
 \caption{Maximum value of mean-square phase fluctuations of \eqref{ph_fluct} 
$\Phi^2 \coloneqq {\rm max} [\langle \hat{\delta \phi}^2_{xx'} \rangle]$
 (peak height in Fig.\,\ref{surf_fluct}), and fragmentation degree, $1 - n_1$ 
 as function of interaction strength $g_0$, for $N=10$. The inset shows the
 ratio of the maximum fluctuation $\Phi^2$ and fragmentation degree $1-n_1$.}
\label{fragm_fluct}
 \end{figure}
The occupation numbers we obtain after convergence of the self-conistent equations has been reached, for $N=10$ and 
varying interaction strength $g_0$, are given in Table~\ref{table10} (for $N=30$ and $N=100$, cf.
the discussion in \,Appendix B). We see that the degree of fragmentation $1-n_1$  
grows rapidly with interaction strength. However, for the $g_0$ ranges we consider the fragmentation degree is still in the range of a few percent only. This, in turn, also 
represents a condition for our calculations, with a fixed number of orbitals $M=5$, to be reliable for the chosen range of 
$g_0$. 

In Fig.\,\ref{surf_fluct}, we display surface plots of the mean-square of quantum phase fluctuations $\langle \hat{\delta \phi}^2_{xx'} \rangle$ 
in the $x$--$x'$ plane for three different values of dimensionless interaction strength. We see two very distinct bulges that emerge 
even for small interaction, which grow in size with increasing interaction strength. The detailed shape and fine structure of the bulges corresponds 
to the shape and weight of the different orbitals in the self-consistent solution for the quantum field. 
The emergence of the bulges has the direct interpretation of the loss of phase coherence between distant parts of the cloud by phase fluctuations. This  phase-phase-correlations induced phenomenon is conjugate to the coherence loss indicated by density-density correlations which was discussed in \cite{Kang}. 

Fig.\,\ref{fragm_fluct} shows the dependence of the maximum value of phase fluctuations $\Phi^2 \coloneqq {\rm max}[\langle \hat {\delta \phi}^2_{xx'} \rangle]$ (measured at the top of the bulges in Fig.\,\ref{surf_fluct}),  
and of the degree of fragmentation $1-n_1$, as functions of the dimensionless interaction strength $g_0$. 
We recognize a very smooth, almost linear dependence of both quantities on $g_0$. 
The inset however also shows that the dependence is more complicated than linear. 
The ratio $\Phi^2 / (1 - n_1)$ can be fitted rather well with an exponential function, which implies a 
complex relation between the maximal fluctuation value and the degree of fragmentation. 
On the other hand, the smooth dependence of $\Phi^2$ 
on $g_0$ we see in Fig.\,\ref{fragm_fluct} 
 is expected from the fact that we observed in our calculations that the shape of the orbitals does not change drastically by increasing the interaction strength. 

In Fig.\,\ref{comparison}, we display the spatial dependence of the mean-square  
phase fluctuations relative to the origin, $\langle \hat{\delta \phi}^2_{x0} \rangle$ 
We observe that the emergence of the local phase-fluctuation maximum we observe using MCTDH in 
Fig.\,\ref{comparison} is akin to the first-order correlation function dips  
obtained in Monte Carlo calculations for the Tonks-Girardeau limit of very strong interactions~\cite{Minguzzi}.
The self-consistent approach MCTDH, similarly, 
therefore contains  many-body correlations between field-operator modes with spatially inhomogeneous modulus,  
which 
local-density approximations, by construction, do not describe.

\begin{figure}[t]
 \center
\includegraphics[width=0.4\textwidth]{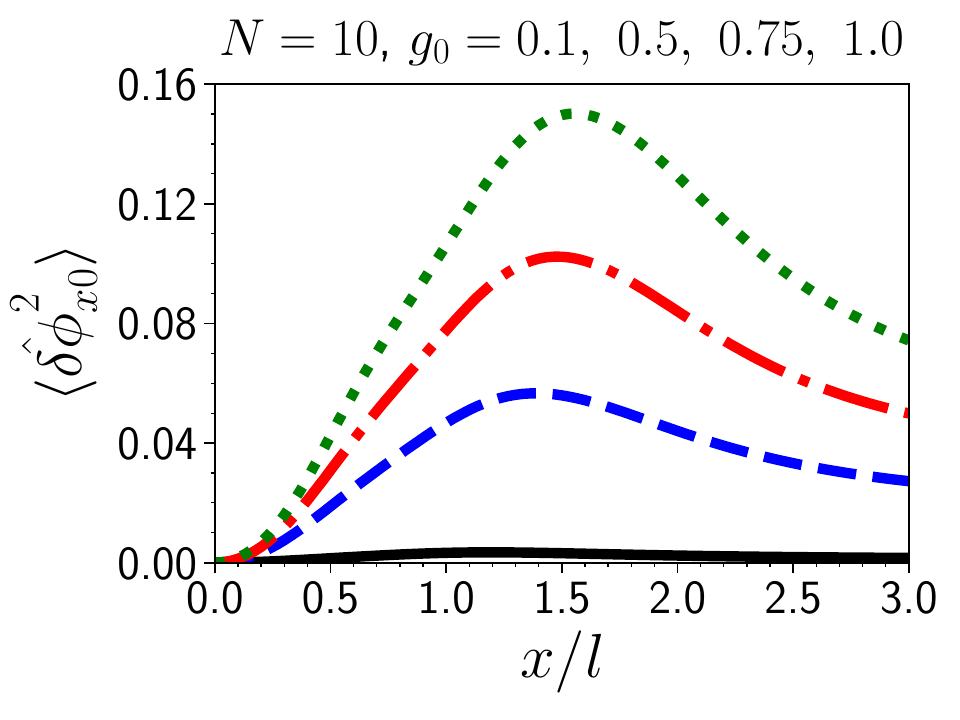}
 \caption{Phase fluctuations $\langle \hat {\delta \phi}^2_{x0} \rangle$
 calculated 
 with MCTDH
 using~\eqref{ph_fluct}, 
 for $N=10$ and varying interaction strength:  $g_0 = 0.1$ (black solid line), $g_0 = 0.5$ (blue dashed line), $g_0 = 0.75$ (red dashed-dotted line), and $g_0 = 1.0$ (green dotted line). 
 We note that $g_0 = 0.1$ shows a similar peak behavior as the other couplings.}
 \label{comparison}
\end{figure}

\section{Discussion}
We have performed self-consistent many-body calculations for a 1D trapped Bose gas with MCTDH-X, which reveal that phase-fluctuating BECs contain many-body correlations between all significantly occupied orbitals, 
necessary to describe the spatial dependence of the phase fluctuations defined in Eq.~\eqref{phase_fluct1}
both qualitatively and quantitatively in the parameter regime under consideration.
Self-consistency in the determination of significantly field operator orbitals and their population statistics is crucial for the accuracy and predictive power of many-body calculations, even when 
the degree of fragmentation 
(defined by the relative number of particles outside the most occupied orbital) 
is as small as on the level of percent.
Future work we envisage includes exploring the relevance of self-consistency to the 
large fragmentation regime, and thus its potential impact on the crossover location 
to the Tonks-Girardeau gas in a single harmonic trap \cite{fang,jacqmin}, {cf.~\cite{Zoellner2}, which has treated with MCTDH this crossover to ``fermionization" (that is, the one-dimensional localization of hardcore bosons), however within a double well.} 

Our results suggest a pathway to implement an experimental benchmark for MCTDH,  
by  verifying its predictive power through experimental means, versus the theoretical arguments put forward, e.g.,  
in \cite{lee,manthe,cosme}, which question the accuracy of the convergence of MCTDH and thus also, ultimately, its validity.	
Further possible extensions in this respect, which probe the latter aspects of convergence to the true many-body solution even more deeply, concern nonequilibrium setups created by starting from phase-fluctuating condensates in the ground state, which are experimentally accessible in current experiments as well \cite{Gring,Langen}. 
We finally note that the primary experimental difficulty for the verification of our predictions lies in extracting sufficiently large correlation signals at the relevant small particle numbers of order $N\sim 10-100$. This should, however, represent no matter of principle obstacle given the rapid
progress in the field of ultracold quantum gases, 
cf.~the extensive correlation function analysis performed in \cite{schweigler2015},  
and future experiments with improved resolution and signal to noise ratio should be able to detect the corresponding weaker signals.

\section{Acknowledgments}
This research was supported by the NRF Korea, Grant Nos. 2014R1A2A2A01006535 and
2017R1A2A2A05001422. OVM also acknowledges support by Grant No. 2015616 of 
the Binational USA-Israel Science Foundation.
{URF thanks J. Schmiedmayer for discussions on the experimental observation 
of our predictions 
during a stay at the TU Vienna. }

\appendix 
\begin{widetext}
\section{Multiconfigurational time-dependent Hartree method}
\label{supp_mctdh}

We present, for the sake of being self-contained, here a brief description of the multiconfigurational time-dependent Hartree (MCTDH) method that we use for our self-consistent many-body calculations.
The method has been generalized to apply for all indistinguishable particles~\cite{mctdhx}; here.
 we focus on bosons. 
For more detailed reviews and examples see Refs.~\cite{Meyer1990,meyer2009,alon2008}. 

A system of $N$ interacting bosons is described by using the time-dependent Schr\"{o}dinger equation
\begin{equation}
 \left [ \sum_{i=1}^{N} \hat h(x_i; t) + \sum_{i>j} \hat W(x_i - x_j) \right ] \Psi(x_1,\dots, x_N; t) = i\hbar \frac{\partial \Psi}{\partial t},
\end{equation}
where $\hat h(x_i; t) = \frac{p_i^2}{2m} + V(x_i)$ is the one-body Hamiltonian, with $m$ as mass of the particles, $x_i$ and $p_i$ as position and momentum operators of a given boson $i$, and $V(x_i)$ as the potential energy. The term $\hat W(x_i - x_j)$ is the pairwise particle interaction operator. The many-body wavefunction, $\Psi(x_1,\dots,x_N; t)$, in the MCTDH formulation is expressed by the following ansatz
\begin{equation}
\label{mbwf}
 \ket\Psi
 = \sum_{\{\overrightarrow{N}\}} C_{\overrightarrow{N}}(t) \ket{\overrightarrow{N}; t},
\end{equation}
where the basis $\ket{\overrightarrow{N}; t}$ consists of all possible symmetrized wavefunction 
products of $N$ particles (\textit{permanents}) distributed over $M$ single-particle functions (\textit{orbitals}), 
where $\overrightarrow{N}=(N_1, N_2, \dots, N_M)$ and $N_1 + N_2 + \dots + N_M = N$, i.e. $N_j$ represents the occupation of 
the orbital $j$, and $C_{\overrightarrow{N}}(t)$ are the time-dependent expansion coefficients. The basis size equals $\frac{(N + M - 1)!}{N! (M-1)!}$, i.e. the total number of distributions of $N$ particles among $M$ orbitals. In the language of second quantization, the permanents can be written as
\begin{equation}
 \ket{\overrightarrow{N}; t} = \frac{1}{\sqrt{N_1!N_2!\dots N_M!}} (b_1^{\dagger}(t))^{N_1}(b_2^{\dagger}(t))^{N_2}\dots(b_M^{\dagger}(t))^{N_M}\ket{\rm vac}.
\end{equation}
The operator $b_j^{\dagger}(t)$ is the time-dependent bosonic creation operator, and $\ket{\rm vac}$ is the vacuum state. Clearly, the ansatz is exact if $M\to\infty$, that is if we consider the full Hilbert space of the problem.
In practice, this is however not possible due to the unavoidable computational constraints for any nontrivial (that is, two-body interacting) problem, however for large enough $M$, i.e., when the occupation of the highest orbitals is negligible, the many-body function~\eqref{mbwf} represents a \textit{numerically exact} solution of time-dependent many-body Schr\"{o}dinger equation.

In order to calculate the expansion coefficients $C_{\overrightarrow{N}}(t)$ and orbitals $\{ \varphi_j(x_i; t),~j=1,\dots, M \}$, one applies the Dirac-Frenkel variational principle to the action functional 
\begin{eqnarray}
S[\{C_{\overrightarrow{N}}(t)}\}, 
\{\varphi_j(x_i; t)\}] 
{&=& 
\int dt \left[  
\bra{\Psi}\hat H - i\hbar\delderiv{}{t} \ket{\Psi} 
- \sum_{j,k=1}^{M} \mu_{jk}(t)(\braket{\varphi_j}{\varphi_k} - \delta_{jk}) \right],
\end{eqnarray} 
where $\mu_{jk}(t)$ are time-dependent Lagrange multipliers, ensuring that the orbitals remain orthonormal. The variational procedure gives the equations of motion
\begin{eqnarray}
\label{mctdh_eqs}
 i\hbar\delderiv{\f C(t)}{t} &=& \f H(t) \f C(t), \nonumber\\
i\hbar \delderiv{\ket{\varphi_j}}{t} &=& \hat P \left [\hat h \ket{\varphi_j} + \sum_{k,s,q,l=1}^M \rho^{-1}_{jk} \rho_{ksql}\hat W_{sl}\ket{\varphi_q}  \right ],
\end{eqnarray}
that can be solved numerically in order to obtain the many-body wavefunction $\Psi(x_1, \dots, x_N; t)$. 
Here, $\f{C}(t)$ is the column vector that consists of all possible expansion coefficients $C_{\overrightarrow{N}}(t)$, 
$\f H(t)$ is a matrix composed of matrix elements of the time-dependent Hamiltonian in the corresponding basis 
$\ket{\overrightarrow{N}; t}$, $\hat W_{sl}=\int\int \mathrm{d} x \mathrm{d}x'\varphi^{\ast}_s (x) W(x-x') \varphi_l(x')$ are the local interaction potentials, $\hat P = 1 - \sum_{k'=1}^M \ket{\varphi_k'}\bra{\varphi_k'}$ is a projection operator, and $\rho_{jk}$ and $\rho_{ksql}$ are the matrix elements of the one-body and 
 two-body density matrices, respectively. 

The system of coupled equations in Eqs.~\eqref{mctdh_eqs} is to be solved both for the orbitals and expansion coefficients together. 
This constitutes the notion of self-consistency in inhomogenous quantum many-body systems of interacting bosons 
we employ throughout the paper. This strongly enhanced degree of self-consistency allows for 
genuine many-body effects to be obtained from our calculations.
\section{MCTDH Results for larger particle numbers}

\begin{table}[hbt]
\begin{tabular}{| l | l | l | l | l | l |}
 \hline
 $g_0$&$n_1$ & $n_2$ & $n_3$ & $n_4$ & $n_5$ \\ \hline \hline
 $0.1$  & $0.99694$ & $0.00220$ & $0.00064$ & $0.00015$ & $0.00006$ \\ \hline
 $0.5$  & $0.97256$ & $0.01633$ & $0.00698$ & $0.00277$ & $0.00136$ \\ \hline
 $0.75$ & $0.95917$ & $0.02303$ & $0.01067$ & $0.00473$ & $0.00239$ \\ \hline
 $1.0$  & $0.94785$ & $0.02843$ & $0.01381$ & $0.00653$ & $0.00337$ \\ \hline
 \end{tabular}
 \caption{Table of relative occupation numbers, $n_i = N_i / N$, for $N=30$, and various 
 dimensionless couplings $g_0$, which are defined in Eq.\,\eqref{g0def} of the main paper.}
\label{table30}
\end{table}

Here, we discuss the results  of our calculations 
with an increased number of particles, $N=30$ and $N=100$. The main features are qualitatively very similar to the $N=10$ case. Tables~\ref{table30} and ~\ref{table100} show the relative
occupation numbers of the orbitals. Note that the energy shift due to the interactions is proportional to the number of particles, i.e. for the same value of the numerical parameter $g_0$ the actual interaction differs for systems with a different number of particles. That explains the slight increase in the fragmentation degree for the systems with larger particle numbers we report below. The degree of fragmentation $1 - n_1$ is again not large, but not negligible even for
relatively weak interaction.

\begin{table}[hbt]
\begin{tabular}{| l | l | l | l | l | l |}
 \hline
 $g_0$&$n_1$ & $n_2$ & $n_3$ & $n_4$ \\ \hline \hline
 $0.1$  & $0.99503$ & $0.00334$ & $0.00113$ & $0.00044$\\ \hline
 $0.5$  & $0.97889$ & $0.01192$ & $0.00622$ & $0.00297$\\ \hline
 $0.75$ & $0.97360$ & $0.01433$ & $0.00808$ & $0.00398$\\ \hline
 $1.0$  & $0.96983$ & $0.01595$ & $0.00946$ & $0.00477$\\ \hline
 \end{tabular}
 \caption{Table of relative occupation numbers, $n_i = N_i / N$, for $N=100$, and various 
 dimensionless couplings $g_0$, which are defined in Eq.\,\eqref{g0def} of the main paper. Note that due to the technical challenges we used $M=4$ number of orbitals for these calculations.}
\label{table100}
\end{table}

In Fig.~\ref{surf_fluct30} we show surface plots of the mean-square of quantum phase fluctuations $\langle \hat{\delta \phi}^2_{xx'} \rangle$ 
in the $x-x'$ plane for various values of the dimensionless interaction strength $g_0$. 
Bulges of a similar geometric shape to the $N=10$ case emerge, cf.~Fig.\,\ref{surf_fluct}, 
since the shape of the correlations is dictated by the geometry of the trapping potential and single-particle orbitals. 
Fig.~\ref{fragm_fluct30} shows the dependence of the maximum value of mean-square phase fluctuations $\Phi^2 \coloneqq {\rm max} [\langle \hat{\delta \phi}^2_{xx'} \rangle]$ and fragmentation degree $1-n_1$, as a function of dimensionless coupling $g_0$. 
We see that $\Phi^2$ grows slightly slower for $N=30$ than for $N=10$, but remains a smooth
and almost linear function of $g_0$, cf.~Fig.~\ref{fragm_fluct}. 
Finally, Fig.~\ref{comparison30} displays 
the spatial behavior of the relative phase fluctuations, again demonstrating the peak 
visible in the  $N=10$ data of~Fig.~\ref{comparison}. 
, 

Table~\ref{tableError30} shows the $N=30$ 
relative approximation errors for energy and occupation numbers for the first and second orbitals, when using $M=5$ and $M=7$ orbitals, respectively.
Analogously to the $N=10$ case shown in Table  \ref{tableError},  
the approximation errors are rather small. Thus we conclude that 
already for $M=5$ orbitals, convergence is achieved.
\begin{table}
\begin{tabular}{| l | l | l | l | l | l |}
 \hline
 $g_0$&$\Delta_{\rm rel}E$ & $\Delta_{\rm rel}n_1$ & $\Delta_{\rm rel}n_2$ \\ \hline \hline
 $0.1$  & $-8.0\times10^{-4}$ & $-1.4\times10^{-4}$ & $2.0\times10^{-2}$ \\ \hline
 $0.5$  & $-6.2\times10^{-3}$ & $-3.3\times10^{-3}$ & $4.8\times10^{-2}$ \\ \hline
 $0.75$ & $-9.3\times10^{-3}$ & $-6.5\times10^{-3}$ & $6.1\times10^{-2}$ \\ \hline
 $1.0$  & $-1.2\times10^{-2}$ & $-1.0\times10^{-2}$ & $7.0\times10^{-2}$ \\ \hline
 \end{tabular}
 \caption{Table of relative approximation errors, $\Delta_{\rm rel} f = (f_{M=7} - f_{M=5}) / f_{M=7}$, calculated using basis sizes $M=5$ and $M=7$ for energy and occupations $n_1$ and $n_2$,
 $f\coloneqq \{E,n_1,n_2\}$, with $N=30$ and various 
couplings $g_0$ [Eq.\,\eqref{g0def}].}
\label{tableError30}
\end{table}

Finally, qualitatively comparable results are also obtained when $N=100$. In Fig~\ref{surf_fluct100} we show surface plots of the mean-square of quantum phase fluctuations as in Figs.~\ref{surf_fluct} and~\ref{surf_fluct30}.. 
The aim here is to demonstrate that the effects described in the main text also appear for a larger (by one order of magnitude), experimentally readily attainable, number of particles. Note that for the Fig.~\ref{surf_fluct100} we used a smaller number of orbitals, $M=4$, due to the increased numerical demand. Note also that the existence of additional local maxima, which can already be observed in Fig.\ref{surf_fluct30}, is more prominent for the larger system $N=100$. 
As in Fig.~\ref{fragm_fluct100}, we plot the maximum value of mean-square phase fluctuations, $\Phi^2 \coloneqq {\rm max} [\langle \hat{\delta \phi}^2_{xx'} \rangle]$, and fragmentation degree, $1-n_1$, as functions of the interaction strength, $g_0$. We can see a tendency to saturation for both, and especially for $\Phi^2$. 

\begin{figure}[hbt]
 \includegraphics[width=0.4\textwidth]{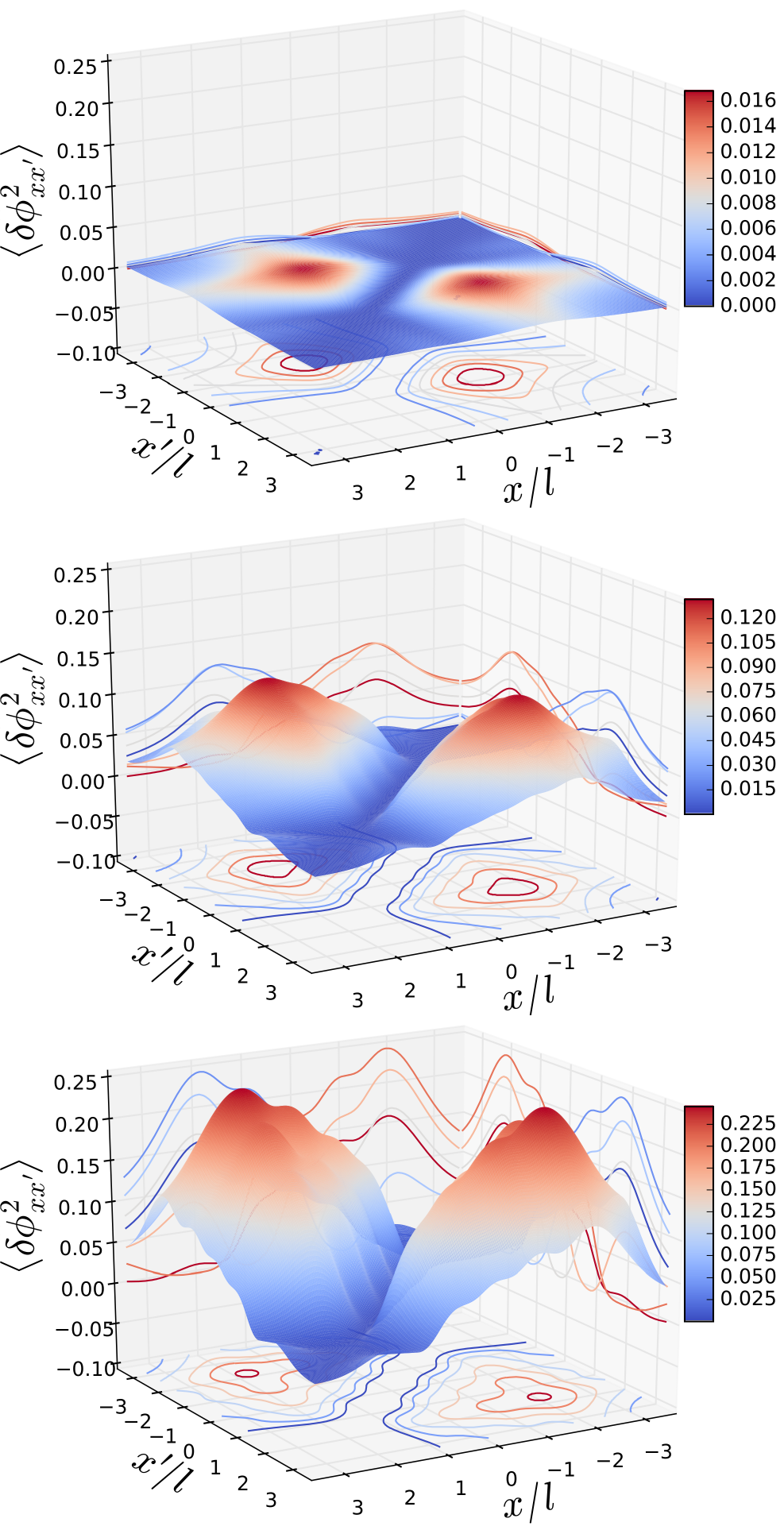}
 \caption{Mean-square quantum phase fluctuations $\langle \hat{\delta \phi}^2_{xx'} \rangle$ for $N=30$ and increasing coupling strength:  
 Top $g_0 = 0.1$, Middle $g_0 = 0.5$, and Bottom $g_0 = 1.0$. 
 The maxima along the off-diagonals, $x'=-x$, correspond to the fact that the gas  
 becomes phase-uncorrelated in distant regions of the cloud.}
  \label{surf_fluct30}
\end{figure}
\begin{figure}[t]
 \includegraphics[width=0.5\textwidth]{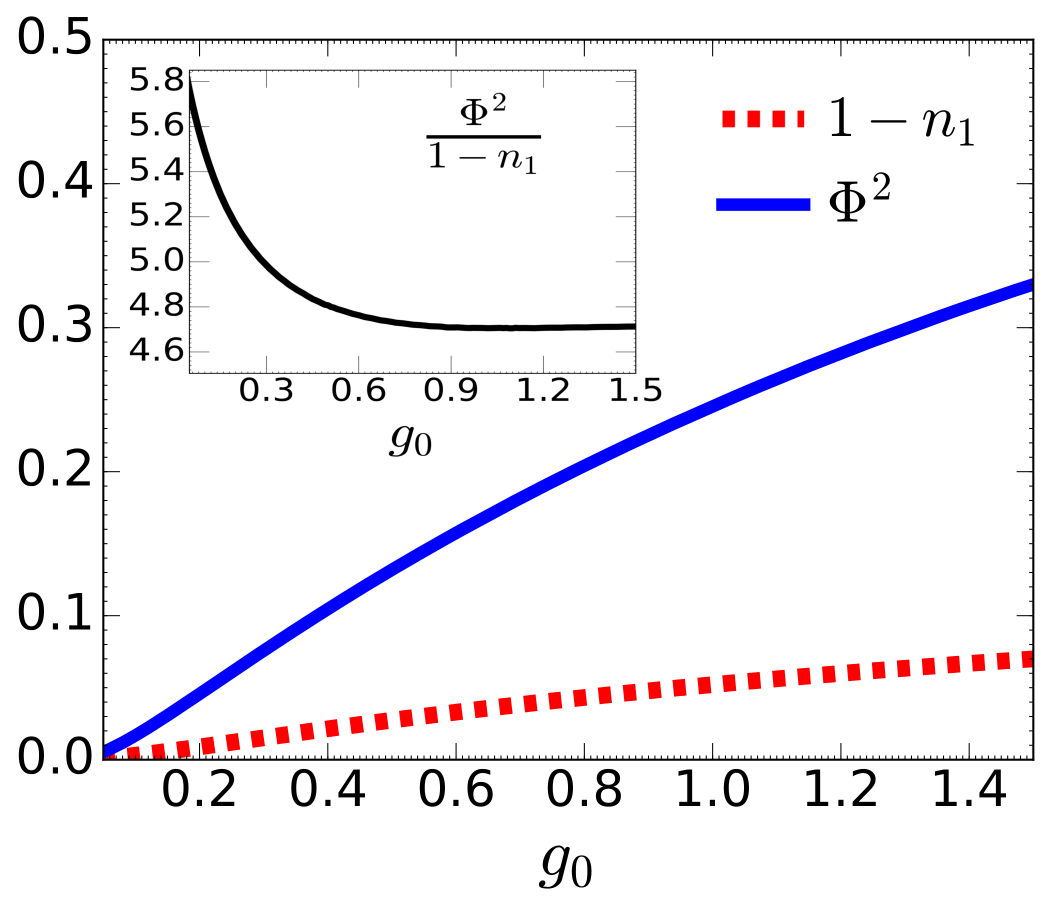}
 \caption{Maximum value of mean-square phase fluctuations of \eqref{ph_fluct} 
$\Phi^2 \coloneqq {\rm max} [\langle \hat{\delta \phi}^2_{xx'} \rangle]$
 (peak height in Fig.\,\ref{surf_fluct30}), and fragmentation degree, $1 - n_1$ 
 as function of interaction strength $g_0$, for $N=30$. The inset shows the
 ratio of the maximum fluctuation $\Phi^2$ and fragmentation degree $1-n_1$.}
\label{fragm_fluct30}
 \end{figure} 
\begin{figure}[b]
 \center
\includegraphics[width=0.4\textwidth]{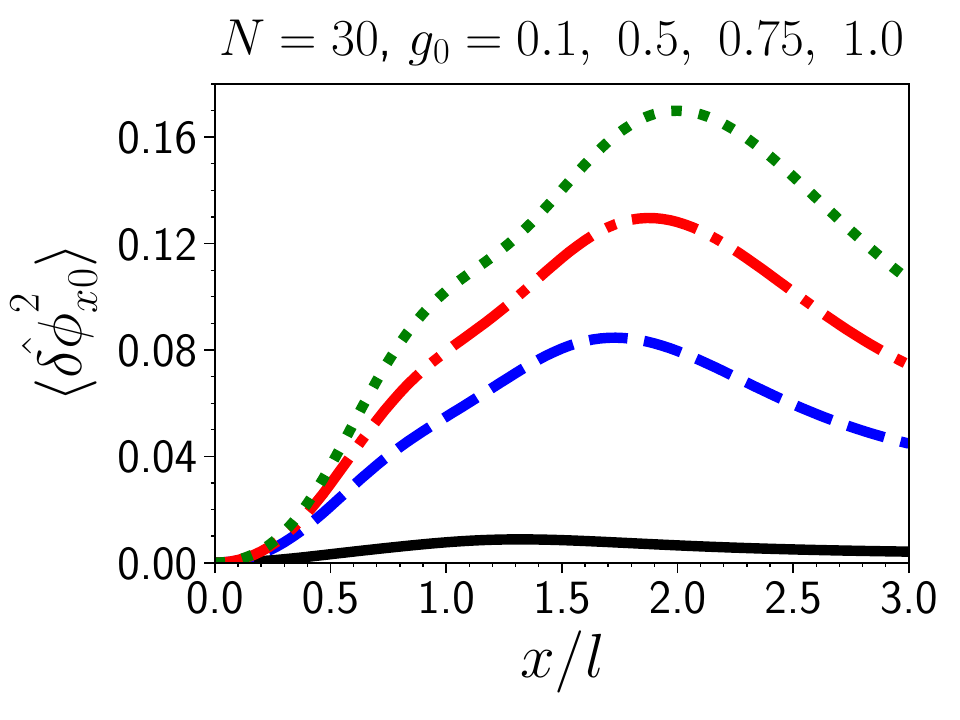}
 \caption{Phase fluctuations $\langle \hat {\delta \phi}^2_{x0} \rangle$,  
 calculated 
 with MCTDH using~\eqref{ph_fluct}, 
 for $N=30$  and varying interaction strength:  $g_0 = 0.1$ (black solid line), $g_0 = 0.5$ (blue dashed line), $g_0 = 0.75$ (red dashed-dotted line), and $g_0 = 1.0$ (green dotted line).}
 \label{comparison30}
\end{figure}
\begin{figure}[hbt]
 \includegraphics[width=0.4\textwidth]{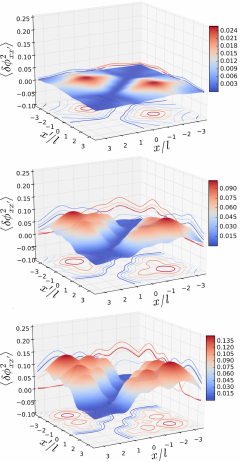}
 \caption{Mean-square quantum phase fluctuations $\langle \hat{\delta \phi}^2_{xx'} \rangle$ for $N=100$ and increasing coupling strength:  
  Top $g_0 = 0.1$, Middle $g_0 = 0.5$, and Bottom $g_0 = 1.0$.   The maxima along the off-diagonals, $x'=-x$, correspond to the fact that the gas  
 becomes phase-uncorrelated in distant regions of the cloud.}
  \label{surf_fluct100}
\end{figure}
\begin{figure}[t]
 \includegraphics[width=0.5\textwidth]{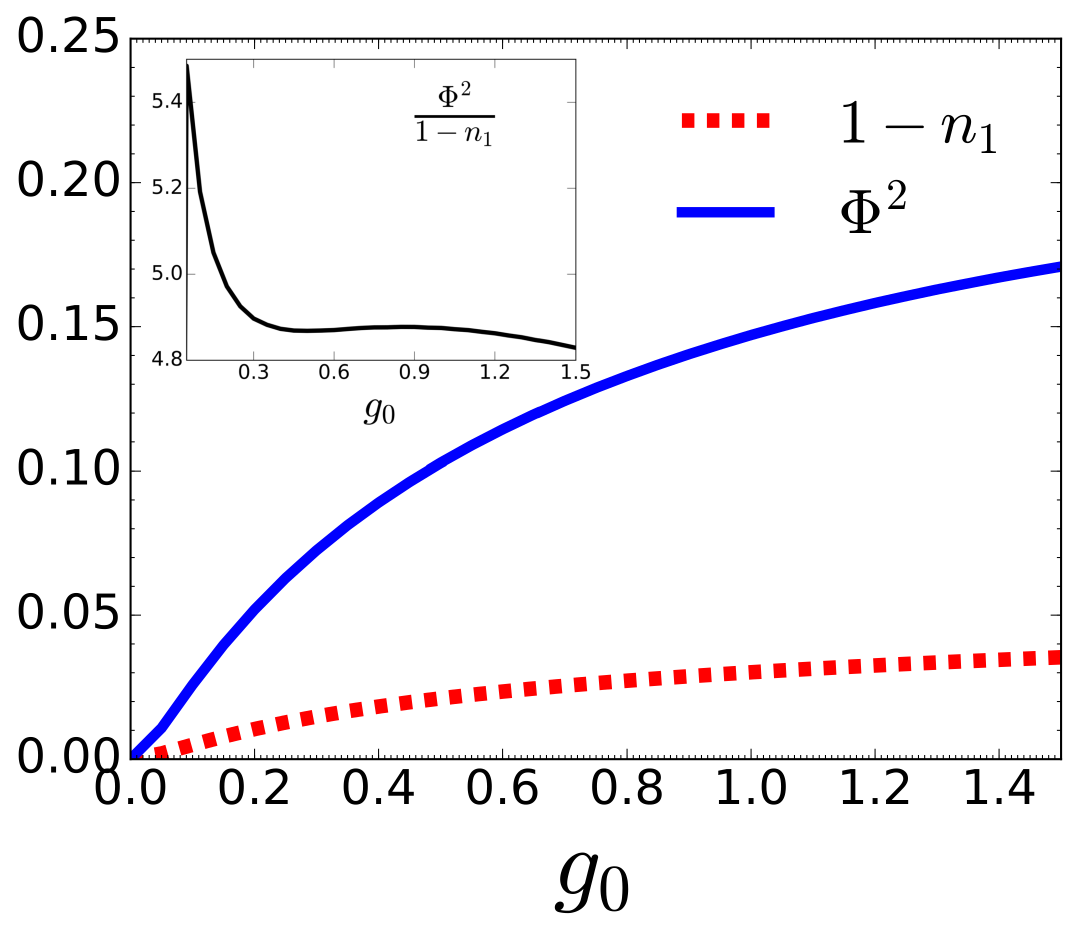}
 \caption{Maximum value of mean-square phase fluctuations of \eqref{ph_fluct} 
$\Phi^2 \coloneqq {\rm max} [\langle \hat{\delta \phi}^2_{xx'} \rangle]$
 (peak height in Fig.\,\ref{surf_fluct100}), and fragmentation degree, $1 - n_1$ 
 as function of interaction strength $g_0$, for $N=100$. The inset shows the
 ratio of the maximum fluctuation $\Phi^2$ and fragmentation degree $1-n_1$.}
\label{fragm_fluct100}
 \end{figure} 
\begin{figure}[b]
 \center
\includegraphics[width=0.4\textwidth]{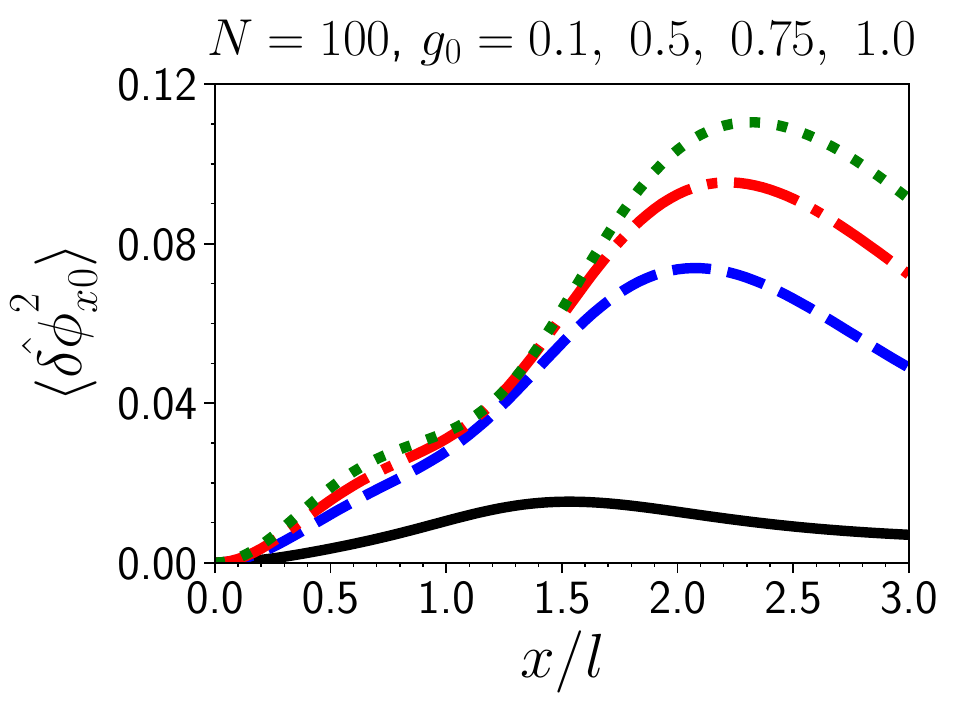}
 \caption{Phase fluctuations $\langle \hat {\delta \phi}^2_{x0} \rangle$,  
 calculated 
with MCTDH  
 using~\eqref{ph_fluct},  
 for $N=100$  and varying interaction strength:  $g_0 = 0.1$ (black solid line), $g_0 = 0.5$ (blue dashed line), $g_0 = 0.75$ (red dashed-dotted line), and $g_0 = 1.0$ (green dotted line). }
 \label{comparison100}
\end{figure}
\end{widetext}

\bibliography{oleks37}

\end{document}